\documentclass[10pt,aps,prx,amsfonts,amsmath,amssymb,raggedbottom,longbibliography,reprint,superscriptaddress,citeautoscript]{revtex4-2}
\usepackage[usenames,dvipsnames]{color}
\usepackage{graphicx,microtype}
\usepackage{float}
\usepackage[bookmarks=false,colorlinks]{hyperref}
\hypersetup{linkcolor=Plum,citecolor=Plum,filecolor=Plum,urlcolor=PineGreen}

\begin{document}

\title{Finite temperature dielectric properties  of KTaO$_3$ from first principles and machine learning: Phonon spectra, Barrett law, strain engineering and electrostriction}

\author{Quintin N. Meier}
\email{quintin.meier@cea.fr}
\affiliation{Université Grenoble Alpes, CEA, LITEN, 17 rue des Martyrs, 38054 Grenoble, France}

\author{Natalio Mingo}
\affiliation{Université Grenoble Alpes, CEA, LITEN, 17 rue des Martyrs, 38054 Grenoble, France}

\author{Ambroise van Roekeghem}
\affiliation{Université Grenoble Alpes, CEA, LITEN, 17 rue des Martyrs, 38054 Grenoble, France}

\date{\today}

\begin{abstract}
Despite important breakthroughs in the last decade, the calculation of temperature dependent properties of solids still remains a challenging task, especially in the vicinity of structural phase transitions. We show that the combination of machine-learning interatomic potentials with quantum self-consistent ab initio lattice dynamics allows to calculate efficiently the temperature dependence of dielectric properties of the quantum paraelectric perovskite KTaO$_3$, with a precision beyond what could be reasonably achieved using plain density functional theory. We first follow the strong anharmonic softening of the polar mode in this incipient ferroelectric material, and the resulting divergence of the dielectric constant that eventually saturates due to the interplay between temperature and quantum fluctuations. Further, we predict the stability range of the quantum paraelectric state under the application of epitaxial strain at 0\ K and 300\ K. Finally, we calculate the temperature dependence of electrostrictive tensors for this material and show that giant electrostriction in KTaO$_3$ is to be expected also at room temperature under the condition of strain engineering.
\end{abstract}

\maketitle
\section{Introduction}
Materials in the vicinity of phase transitions show a divergence in their response functions, making them highly susceptible to external stimuli like electric fields or stresses.  In the case of a ferroelectric phase transition, the diverging response function is the dielectric constant, which follows a Curie-Weiss-law $\epsilon \propto \frac{1}{T-T_C}$ \cite{devonshire1954}, where $T_C$ is the critical temperature. The divergence of the dielectric constant is rooted in the softening of an optical phonon branch. The frequency of this branch is expected to follow a power-law behavior $\omega\propto(T-T_C)^{\frac{1}{2}}$. At $T_C$, a spontaneous polarization starts developing and the dielectric constant starts decreasing again. In some materials, critical softening of a optical phonon mode is observed down to very low temperatures, but the phase transition is never reached. In these incipient ferroelectrics the zero-point motion of the atoms is stabilizing the paraelectric state even at 0~K, which is why they are often referred to as quantum paraelectric. Close to 0~K, when temperature fluctuations freeze out, the dielectric constant deviates from the the Curie-Weiss behavior and flattens out. This low-temperature behavior was first described by Barrett \cite{Barrett1952}, and more accurate low-temperature scaling laws were developed later \cite{Schneider1976,oppermann1975critical,Vendik1997, rowley2014ferroelectric}. 

A typical example of a quantum paraelectric material is the perovskite KTaO$_3$, which shows a strong increase in the dielectric constant at low temperatures, which eventually saturates between 5 and 15\ K to a zero-point value of $\approx$ 4500 \cite{wemple1965,ang_2001,fujishita2016}. The associated soft optical phonon frequency has also been measured and flattens out at $\approx$ 15\ K \cite{shirane1967}.

The goal of this work is to provide a accurate description of the dielectric properties of KTaO$_3$ fully on the basis on Density Functional Theory (DFT) calculations. DFT is the established tool in the calculation of phonon band structures, which can be obtained via the calculation of the force constants matrix ($\Phi_{ij}=\dfrac{\partial^2 E}{\partial u_i \partial u_j}$), which is the second derivative of the total energy ($E$) with respect to atomic displacements ($u_i$). DFT has also become an essential tool to study changes of the crystal structure at structural phase transitions, as it allows to compute the energy landscape of different distortion modes. However, one of the big limitations for DFT is the fact that the energy landscape is obtained only in the 0\ K limit and without including neither temperature dependent fluctuations nor zero-point motion of the atoms. One is therefore often limited to use DFT-parametrized phenomenological theories and effective Hamiltonians to calculate temperature dependent phonon spectra and dielectric constants, see e.g. \cite{akbarzadeh2004,meier2020,esswein2021ferroelectric, Shin2021,meier2022}, However, obtaining quantitative agreement with experimental data has been a challenge due to the large spectrum of anharmonic interactions to be considered.
This has been made possible by the development of new methods to calculate temperature dependent phonon anharmonicities, starting from the work of Souvatzis \textit{et al.} \cite{souvatzis2008entropy}, who took into account classically averaged thermal displacements to self-consistently compute phonon frequencies at high temperatures (SCAILD). Subsequently, we can cite among others the development of the Temperature Dependent Effective Potential (TDEP) \cite{hellmann2011} method, based on a fit of the force constants using ab initio molecular dynamics data, the Stochastic Self-Consistent Harmonic Approximation (SSCHA) \cite{Errea,monacelli2021stochastic}, which is a stochastic approach to find force constants that minimize the harmonic free energy as a function of the density matrix, or self-consistent phonon theory (SCPH) \cite{Tadano2015}, which implements a Green's function approach to calculate anharmonic interactions.

Here, we rely on Quantum Self-Consistent Ab Initio Lattice Dynamics (QSCAILD) \cite{Ambroise_ScF3,qscaild}, which uses a comparable self-consistency constraint as in the original SCAILD, however including the quantum statistics of phonons. 
The methods discussed above rely on a sampling of the energy landscape, which, if done using DFT, leads to a very large computational cost -- typically one order of magnitude higher than small displacements for a single temperature calculation. In order to decrease the computational effort, we take advantage of the progress in machine-learning interatomic potentials (MLIP), which are classical interatomic potentials that are trained using DFT data \cite{behler}. This methodology has a strong added value for self-consistent methods, since it allows to massively increase the amount of configurations in the stochastic sampling. Conversely, since for the accuracy of the stochastic method it is essential to create a number of uncorrelated configurations, this provides a very effective training set for the machine-learned interatomic potential.

We demonstrate the advantages of this methodology on the case of KTaO$_3$ close to the quantum critical point, showing that it captures the saturation of the soft mode and dielectric constant following Barret's law. We then explore the stability of the paraelectric state under changes of the volume due to isotropic strain as well as epitaxial strain, both at 0\ K and 300\ K. Recently, the potentially giant electrostriction in this material has caught a renewed attention \cite{tanner,yu2022}. We confirm these previous reports and further calculate the temperature dependence of the electrostrictive coefficients. We obtain a constant value of the strain response as a function of polarization, showing that giant electrostriction in KTaO$_3$ should also be expected at room temperature with proper strain engineering.

\section{Methods}
\subsection{Lattice dynamics including quantum and thermal fluctuations}
Typical calculations of interatomic force constants in real space are done based on the computation of forces due to small ionic displacements from the DFT ground state. To take into account the quantum and thermal fluctuations of the atomic structure, we rely on a self-consistent sampling of atomic configurations in a harmonic model of the quantum canonical ensemble, coupled with a regression analysis of the forces. In this scheme, the probability $\rho(u)$ of finding the atoms in a supercell displaced by a vector $u$ is proportional to $\exp\left[-\dfrac{1}{2}u^T\Sigma^{-1} u\right]$, where $\Sigma$ is the quantum covariance matrix, which is given by
\begin{equation*}
    \Sigma_{i\alpha,j\beta}=\frac{\hbar}{2\sqrt{M_{i}M_{j}}}\sum_{m}\omega_{m}^{-1}\left(1+2n_{B}\left(\omega_{m}; T\right)\right)\epsilon_{mi\alpha}\epsilon_{mj\beta}^{*}
\end{equation*}
with $n_B$ the Bose-Einstein distribution, $M$ the masses of atoms, $\omega$ the phonon eigenvalues and $\epsilon$ the eigenvectors. Starting from an initial set of force constants, a series of atomic configurations is generated, and using a least-square regression of the computed forces we obtain a new set of effective force-constants. Self-consistency is obtained by iteration, if the phonon spectrum is stable. The volume is optimized by taking into account both the potential stresses directly obtained from the DFT calculation as well as the kinetic stress, that is obtained using the quantum virial theorem \cite{fock1930}.
In practice, our method is implemented in the Quantum Self-Consistent Ab-Initio Lattice Dynamics (QSCAILD) package, which is described in further detail in Ref.\ [\onlinecite{qscaild}].

\subsection{Density functional theory calculations}
For our density functional theory calculations we use DFT as implemented in the Vienna Ab Initio Simulation package (VASP) version 5.4.4 \cite{vasp1,vasp2,vasp3,vasp4}. We use the PBEsol functional with an energy cutoff of 600 eV. For the calculation of the Born effective charges we use a wavevector grid of $8\times 8 \times 8$. For the calculations of different atomic configurations used in QSCAILD we use a 4x4x4 unit cell (392 atoms), using a wavevector grid of $1\times 1\times 1$.

\subsection{Machine-learning inter-atomic potentials}
\begin{figure}[t]
    \centering
    \includegraphics[width=0.4\textwidth]{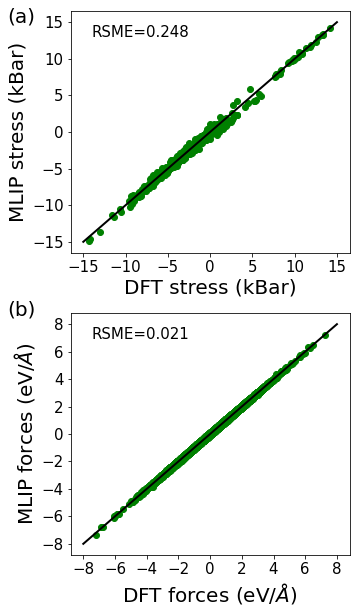}
    \caption{(a) Stresses and (b) forces for a total of 160 test (80 at 0\ K and 80 at 300\ K) configurations as a comparison between DFT and predictions from the machine learnt potential.}
    \label{fig:stress_force}
\end{figure}
Using QSCAILD, our main bottleneck is the number of DFT calculations required to obtain stochastic convergence, which typically leads to more than one order-of-magnitude increase in computational effort compared to a simple phonon calculation using small displacements. To bypass this difficulty, we have developed an interface between the QSCAILD code and moment tensor potentials as implemented in the MLIP package \cite{Novikov_2021}. We use the implemented active learning scheme on the sampled configurations, which allows for on-the-fly training of the inter-atomic potential. Potentials with 16 moment tensor descriptors (see \cite{Novikov_2021}) were used in the machine learning part. The active learning algorithm is governed by the calculation of the extrapolation grade \cite{Novikov_2021}.

This leads to a substantial benefit since previous DFT calculations -- from a previous iteration cycle, a different temperature or pressure -- can now be used to train the interatomic potential. In a typical study including several temperatures, this can lead to a speedup of 2 orders of magnitude or more. Moreover, since the overhead to use additional atomic configurations becomes negligible, this allows to reduce the stochastic noise inherent to this method and thus brings new possibilities such as exploring smaller temperature variations, or conducting a more complete study including both temperature and pressure, as shown in this manuscript. We note that reweighting schemes were already used previously to allow for reusability of the DFT calculations \cite{Errea,qscaild}. However, this reweighting scheme is not transferable to different lattice parameters, and can lead to biased estimations, while the machine-learnt interatomic potential with active learning is designed to handle such difficulties. During the training of the MLIP, we used equal weights of 1.0 for the training of stresses and forces, and a low weight for the total energy (0.01), as the total energy is not used in the QSCAILD method. The excellent agreement between the stresses and forces predicted by the machine-learnt potential and those from DFT are shown in Fig. \ref{fig:stress_force}, on a test set of 160 configurations from QSCAILD calculations at 0\ K and 300\ K.

\subsection{Calculation of the static dielectric constants}\label{sec:dec}
The static dielectric constant can be calculated directly from the force constants matrix that is calculated with the QSCAILD methodology in combination with the Born effective charges $Z^*$ (BECs) calculated from density functional perturbation theory \cite{Gonze1997, Ambroise_dielectric}. From the diagonalized force constant matrix we obtain the eigenvectors $\xi^n$ and the eigenvalues $a_n$.

Each eigenmode $n$ with a non-zero mode-effective charge $Z_{i}^n=\sum\limits_{j} Z^*_{ij}\xi^n_j$, will contribute to the total susceptibility $\chi_{ij}$ by a total amount of

\begin{equation}
    \chi^{n}_{ij}=\dfrac{1}{\epsilon_0}\dfrac{Z_{i}^nZ_{j}^n}{a_n\Omega}
\end{equation}
where $Z_{i}^n=\sum\limits_{j} Z^*_{ij}\xi^n_j$ is the mode effective charge of mode $n$ along direction $i$. Finally the total dielectric constant is given by the sum of the responses for all the eigenmodes,

\begin{equation}
    \epsilon=1+\chi_{ij}^\infty+\sum\limits_{n}\chi_{ij}^n \quad 
\end{equation}
where $\chi_{ij}^{\infty}$  is the electronic response, and $\chi^n_{ij}$ is the ionic response for each eigenmode $n$ of the force constant matrix.

\section{Results}
\subsection{Temperature-dependent phonon spectrum, dielectric constant and Barrett's law}
\begin{figure}[b]
    \centering
    \includegraphics[width=0.5\textwidth]{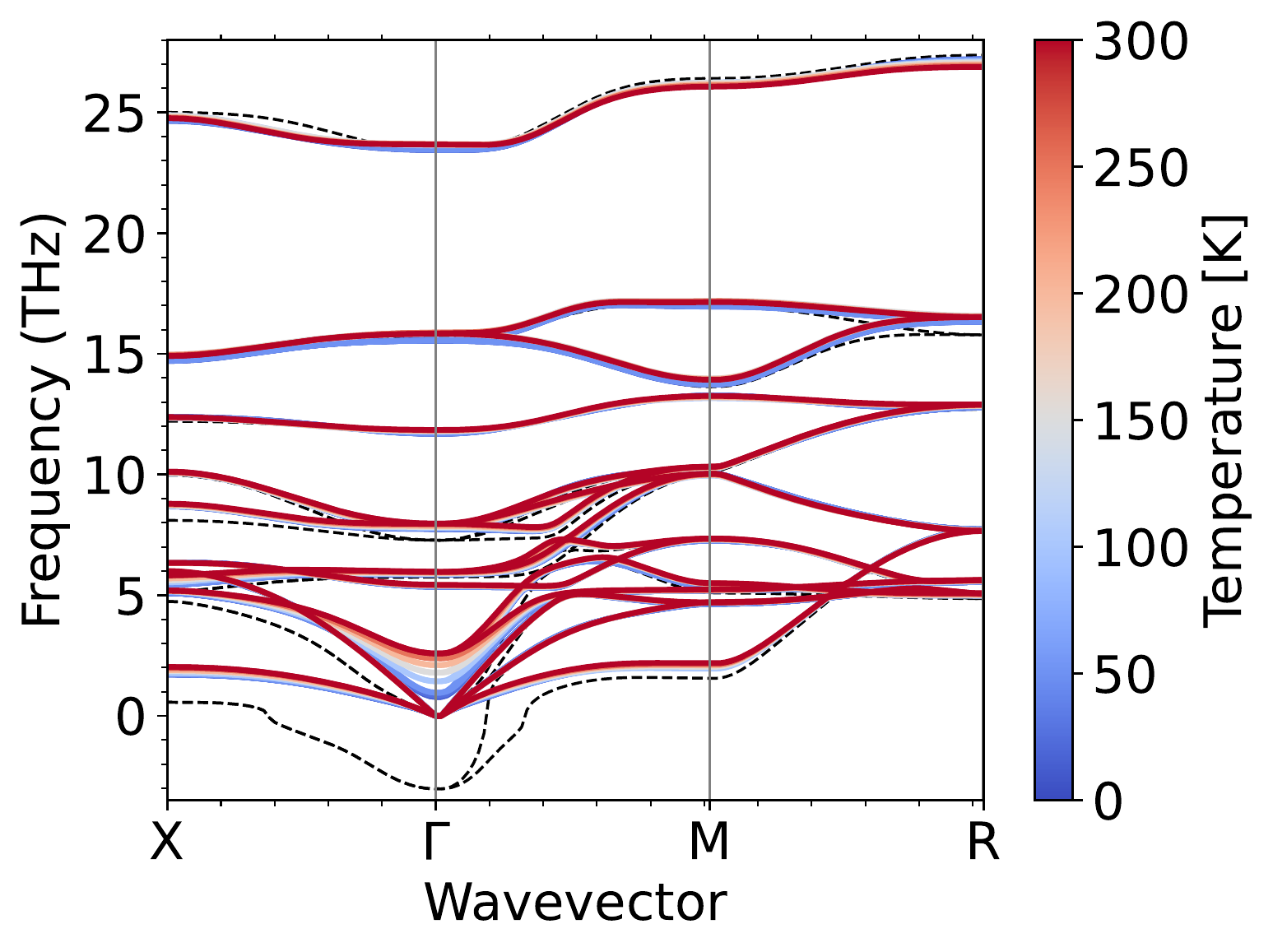}
    \caption{Temperature dependent phonon spectrum of KTaO$_3$. The color lines show the evolution of the bands with temperature, and the black dashed lines shows the bands calculated with small displacements.}
    \label{fig:spectrum_KTO}
\end{figure}
\begin{figure}[h]
    \centering
    \includegraphics[width=0.4\textwidth]{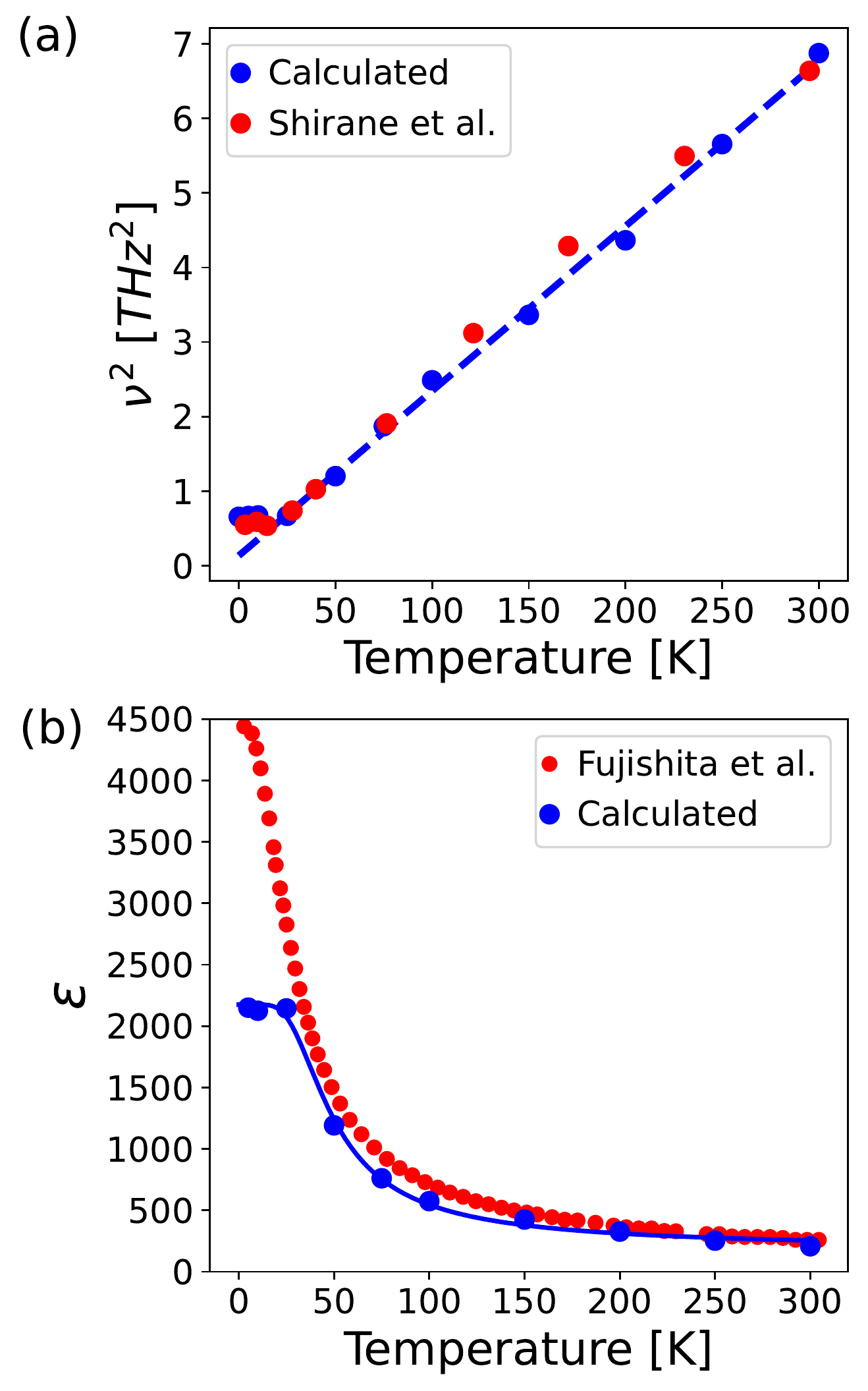}
    \caption{(a) Squared frequency of the soft mode, our calculations compared to the frequencies measured by Shirane \textit{et al.} [\onlinecite{shirane1967}] (b) Calculated dielectric constant compared to the measurements by Fujishita \textit{et al.} [\onlinecite{fujishita2016}]. The fitted line corresponds to a fit to the function derived by Barrett \cite{Barrett1952}.}
    \label{fig:soft_mode}
\end{figure}

We start by calculating the phonon spectrum of KTaO$_3$ using density functional theory (DFT) and small displacements. We find a mode with imaginary frequency at the $\Gamma$ point (See Figure \ref{fig:spectrum_KTO}).
Modes with imaginary frequencies correspond to a negative second derivative of the force constant matrix, thus the mode is energy lowering. The observed mode corresponds to an irreducible representation with $\Gamma_4^-$ symmetry, and it corresponds to a polar soft mode, very much alike the ferroelectric softmode in other perovskites (like PbTiO$_3$, BaTiO$_3$, BiFeO$_3$,..). This polar mode is the one dominating the low-temperature dynamical and dielectric properties in this system.  Experimentally, it is know that KTaO$_3$  remains cubic at 0\ K due to its quantum paraelectric nature. In order to take into account the zero-point motion of the atoms, we perform a QSCAILD calculation at 0\ K. In the converged spectrum, we find that even at 0\ K, the unstable mode remains stable, in accordance with experimental observations. We proceed to calculate the temperature dependence of the phonon spectrum for a series of temperatures (see Figure \ref{fig:spectrum_KTO}). We observe that the temperature dependence of the phonon spectrum is mostly concentrated on the ferroelectric soft mode which rises from 0.8\ THz at 15\ K to a frequency of 2.6\ THz at 300\ K. This is in excellent agreement with experimental data \cite{Perry1989, shirane1967}, which show a temperature evolution of the soft mode frequency from $\approx$0.75\ THz at 20\ K, to $\approx$2.7\ THz at 300\ K.
We observe that the frequencies of the soft mode (see Figure \ref{fig:soft_mode}a)) follow a typical $\omega^2\propto T$ behavior above approximately 25\ K, below which we observe a flattening out. This flattening out is consistent with observations in experiments and can be interpreted as the onset of a Barrett-type law \cite{Barrett1952}. Note that from a practical point of view, the machine-learnt potential was crucial for the success of such a calculation, as the lower stochastic sampling would lead to typical error bars of the order of 0.5 THz$^2$ \cite{qscaild}.

Because of the low frequency of this observed polar soft mode, the dielectric constant will be dominated by its ionic contribution. Thus, the temperature dependence of the mode should be reflected in the value of the dielectric constant. We calculate the static dielectric constant using the method described in section \ref{sec:dec} and find good agreement with experimental measurements, with the dielectric constant decreasing from around $\approx$2300 at 0\ K to $\approx$300 at 300\ K. In figure \ref{fig:soft_mode}b) we show our calculations in comparison with experimental data from Ref.\ \cite{fujishita2016}.

When the frequency of the soft mode stabilizes below 25\ K, we observe the same for the dielectric constant. This makes sense physically, since at this frequency the soft mode becomes populated only by the zero-point motion of the material, since $h\nu \approx k_BT$ at around 36~K, which means that thermally induced occupations of the phonon bands freeze out and the dielectric constant reaches the quantum limit.

In experiment however, the dielectric constant keeps rising until 10\ K, where it stabilizes at a value of $\approx$4500. This additional lowering of the frequency below 15\ K is neither visible in the experimental data of Shirane et al \cite{shirane1967}, nor in our calculations.  This might be just an effect of the error bars of the obtained frequencies in both measurements and calculations. As the dielectric constant is proportional to $1/(\nu^2)$, small variations of already low frequencies inevitably lead to large variations in the dielectric constant. In fact, for such low frequencies, an error bar on the order of 0.1 THz on the calculated phonon frequencies already corresponds to an error of around a factor two for the dielectric constant. Additionally, critical fluctuations at the origin of local correlations and structural disorder of the polarization, but also the presence of defects, could induce the polar nanoregions that have been reported in the literature \cite{akbarzadeh2004,Aktas2014, trybula2015}. Such effects are naturally not captured by our harmonic model, and would lead to an additional increase of the dielectric constant in this low-temperature region.

\subsection{Quantum critical points}

In the previous section we have shown that our approach of QSCAILD combined with machine-learnt interatomic potentials is able to quite accurately reproduce the quantum behavior of the soft mode frequency and of the dielectric constant at ambient pressure and low temperatures, despite some discrepancies due to the high sensitivity of the dielectric constant close to the quantum critical point. We will now explore the pressure- and temperature- dependent behavior of the dielectric properties around this critical point, towards the transition to a ferroelectric phase, which can be introduced by both by the application of pressure \cite{uwe1975} and epitaxial strain \cite{tyunina2010,tanner,esswein2021ferroelectric}.

\subsubsection{Volume and pressure}

\begin{figure}[h]
    \centering
    \includegraphics[width=0.45\textwidth]{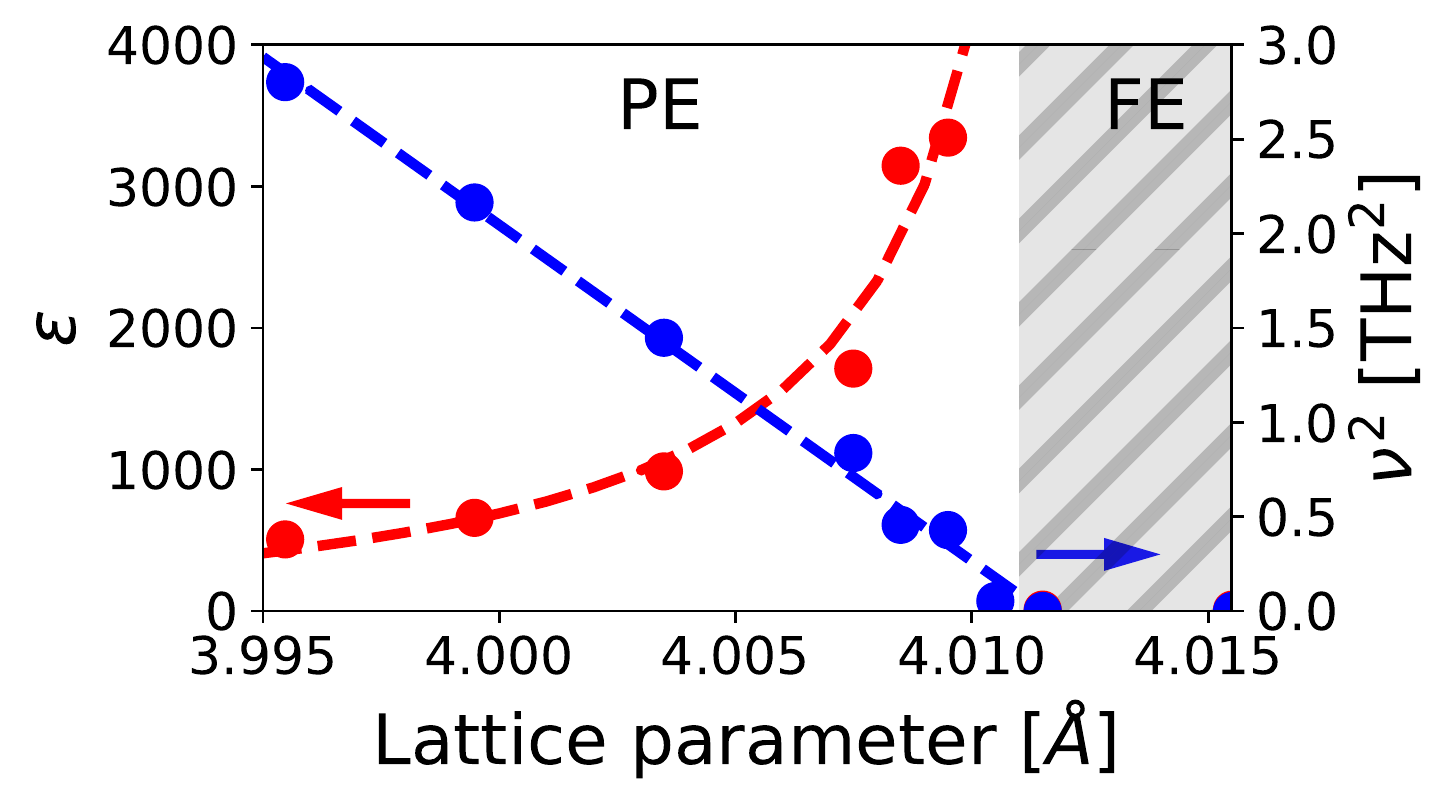}
    \caption{Dielectric constant and squared frequency of the soft mode of KTaO$_3$ at 0\ K as a function of the cubic lattice parameter. The lines correspond to the fit to a Curie-type law for $\epsilon$ and to a fit to $\nu^2\propto T$.  }
    \label{fig:volume}
\end{figure}

The frequency of the ferroelectric soft mode in KTaO$_3$ strongly depends on the lattice parameters. In order to quantify this effect, we calculate the frequency of the soft mode at 0\ K as a function of volume. In our calculations we obtain a good agreement of the lattice constant of 4.007 (at 0\ K), compared to the experimental value of 3.990 by Vousden \textit{et al} \cite{Vousden1951}. Varying the volume at 0\ K, we see that an increase of the volume leads to a power-law decrease in mode frequency and an increase in the dielectric constant. We find again a Curie-type law for the frequency as a function of volume, showing that $\omega^2\propto(a-a_C)$ and $\epsilon\propto \frac{1}{a-a_C}$ (See Figure \ref{fig:volume}).
We find that already an increase of 0.5 \% from our equilibrium lattice parameter leads to a imaginary soft mode frequency and thus the phase transition into  a ferroelectric phase (See Fig. \ref{fig:volume}). The ferroelectric quantum phase transition is in this case induced by an increase of the lattice parameter, and thus effectively by negative isostatic strain, which, in theory, could be induced by doping or vacancy formation \cite{xi2017}. However, an easier way to tune the system towards quantum criticality is the application of epitaxial strain, which is a common method to induce ferroelectricity and control the polar anisotropy axes in ferroelectrics. In this section we will next study the stability of the ferroelectric phase under the application of epitaxial strain both at 0\ K and at finite temperature.

\subsubsection{Epitaxial strain}

\begin{figure*}[htb!]
    \centering
    \includegraphics[width=0.9\textwidth]{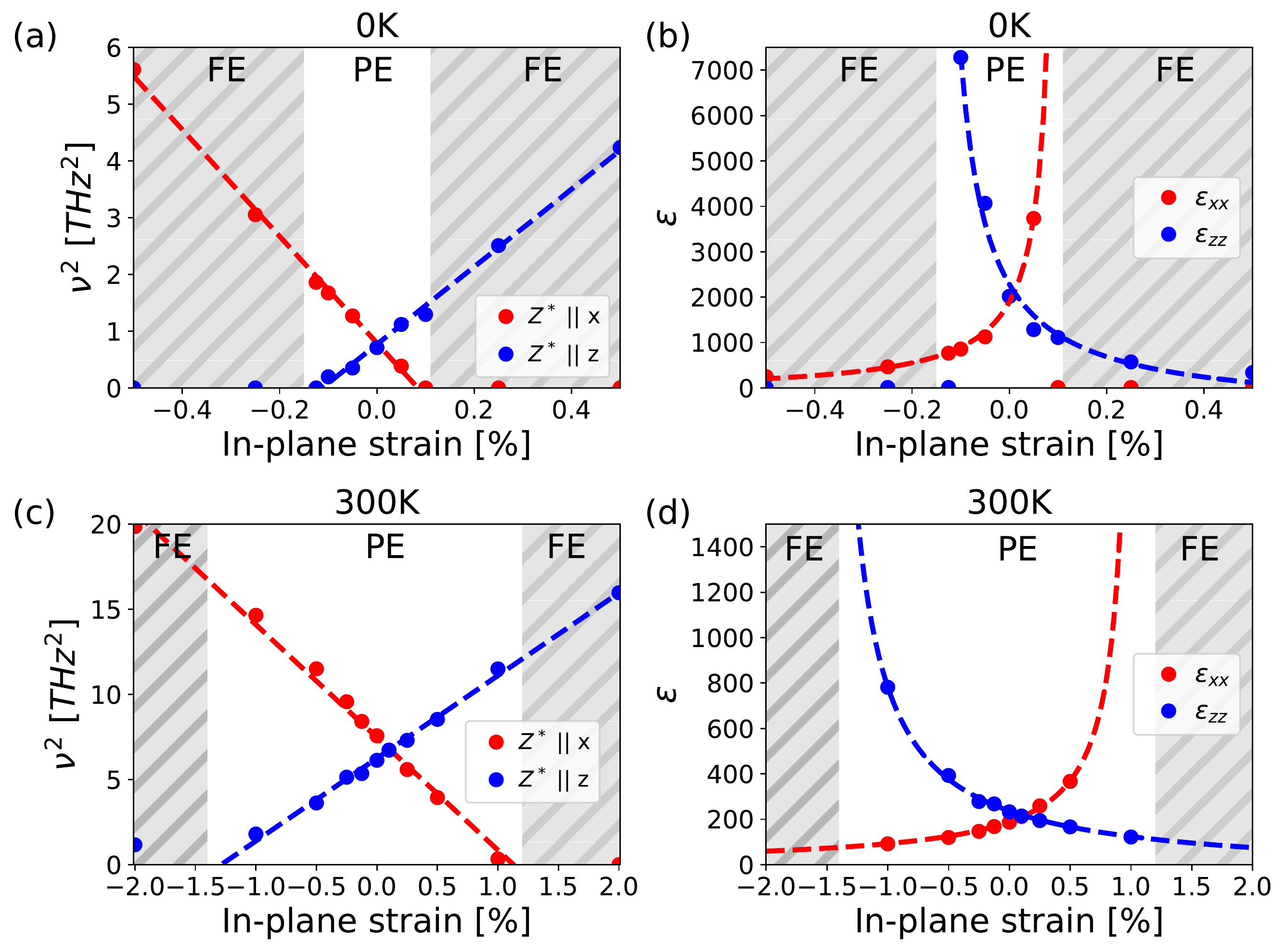}
    \caption{Squared frequencies and dielectric constants under the influence of in-plane epitaxial strain at 0\ K (a,b) and 300\ K (c,d). In the dashed grey areas we observe an unstable phonon, thus in these areas the paraelectric (PE) state is unstable and a ferroelectric state (FE) becomes favorable. $Z^* || x$ describes the doubly degenerate phonon branch with in-plane polarization, while $Z^* || z$  describes the phonon mode with out-of-plane polarization. The plotted lines correspond to linear fits of the frequency$^2$ to the strain, and of Curie-type fits of the dielectric constant.}
    \label{fig:epitaxi}
\end{figure*}

Epitaxial strain has been a widely used tool to control the ferroelectric polarization in thin films of materials. It allows to control both polarization and anisotropy axes in thin film ferroelectrics \cite{ederer2005}. In KTaO$_3$, epitaxial strain has been predicted from first principles calculations to induce a ferroelectric phase transition \cite{tyunina2010,tanner}, however these studies were done without considering the influence of temperature. The application of in-plane strain breaks the cubic symmetry of bulk KTaO$_3$, and thus leads to a splitting of the three-fold degenerate soft mode branches at $\Gamma$, to a doubly degenerate branch in-plane with in-plane polarizations, and a branch with out-of-plane polarization. The squared frequencies of both branches are plotted in Fig \ref{fig:epitaxi}a) and \ref{fig:epitaxi}c). As a result also the degeneracy of the out-of-plane and in-plane dielectric constants is broken. In the presence of compressive strain, the frequency of the out-of-plane polarized phonon decreases, leading to an increase of the out-of-plane dielectric constant. Eventually the out-of-plane phonon becomes unstable and a phase transition into a uniaxial out-of-plane ferroelectric phase is induced.  For tensile strain, the effect is opposite: elongation of the lattice parameters along in the plane leads to a decrease in frequency for the in-plane polarized phonon branches, and thus to a divergence of the in-plane dielectric constant in the plane ($\epsilon_{xx}$=$\epsilon_{yy}$), while the out-of-plane dielectric constant is becoming smaller. Further increase of the tensile strain eventually leads to a phase transition into an in-plane ferroelectric phase (Fig \ref{fig:epitaxi}).
As the dielectric constant is large at 0\ K and the frequency is low, the material is extremely close to a ferroelectric phase transition. We find that the stability range of the paraelectric phase at 0\ K is indeed extremely narrow, spanning from around -0.15 \% compressive to around 0.1\% tensile strain (See Fig \ref{fig:epitaxi}a and b). This is slightly smaller than the values for the stability range reported from ground state DFT calculations, where Tyunina \textit{et al.} report 2 \% \cite{tyunina2010} using LDA  and  Tanner \textit{et al}\cite{tanner} 0.4\%  with PBEsol.  This is however consistent with the fact that we obtain a slightly larger lattice parameter and slightly lower frequencies compared to these works.
At 300\ K, KTaO$_3$ is much further away from the phase transition, and thus the paraelectric state is more resilient to the application of strain.  We find that at 300~K, the paraelectric state is stable in a range of  -1.5\% to 1\% in-plane strain (See figure \ref{fig:epitaxi}c and \ref{fig:epitaxi}d). In addition to the  frequency of the soft mode in the unstrained material being much higher at 300~K, we also observe a difference in slope of the $\nu^2$ lines, which are reduced by around 30\% at 300\ K  compared to the 0\ K calculations. Thus, the strain dependence of the soft mode at finite temperature can not easily be extrapolated from the ground state DFT results.

\subsection{Electrostriction}
In the previous section we have shown that the dielectric constant depends strongly on the volume and the tetragonal strain state of the material. Naturally, this should imply that the application of a small electric field leads to a large polarization as well as strain response. Since in paraelectric KTaO$_3$ piezoelectricity is symmetry-forbidden, the lowest allowed strain response is given by the electrostriction. While piezoelectricity describes a linear coupling between strain and electric field, electrostriction couples the strain to the square of the electric field. This coupling is independent on the space group symmetry and as thus electrostriction is an inherent property of all dielectric materials. It is described by the two second-rank electrostrictive tensors: $Q_{ikl}$ and $M_{ikl}$, where the former describes the strain response to an imposed polarization and the latter describes the strain response to an applied electric field.

\begin{figure}[h]
    \centering
    \includegraphics[width=0.45\textwidth]{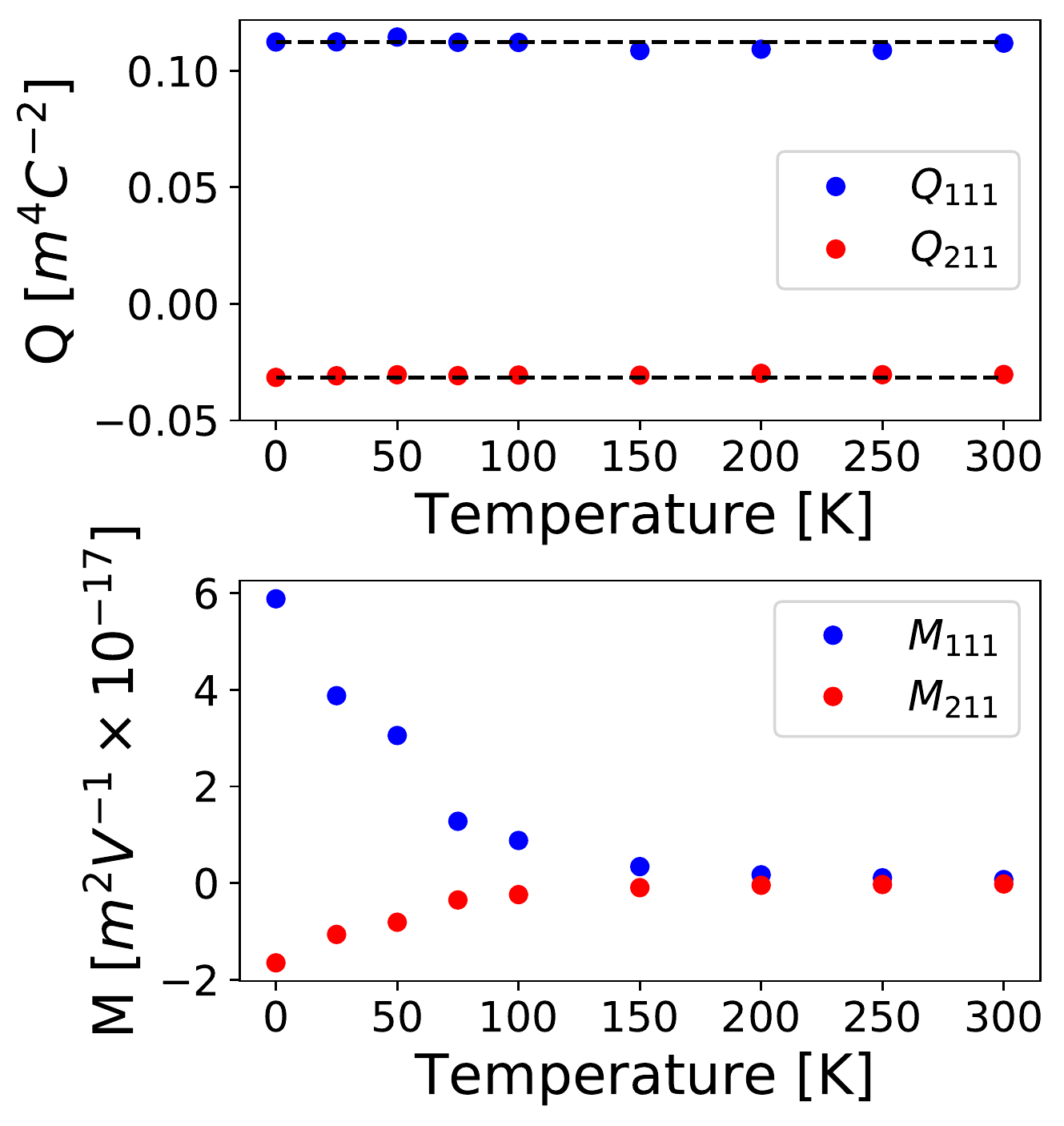}
    \caption{Finite temperature electrostrictive tensors Q (a) and M (b) calculated using QSCAILD with MLIPs.}
    \label{fig:electrostriction}
\end{figure}

The electrostrictive tensors are defined as \cite{devonshire1954}:
\begin{align}
    x_{i}&=Q_{ikl}P_{k}P_{l}\quad ,\\
    x_{i}&=M_{ikl}E_{i}E_{k} \quad, 
\end{align}
where $x_{i}$ is a component of the strain tensor in Voigt notation. In practice, we can calculate the electrostrictive tensors as the second derivative of the strain with respect to a manually imposed polarization along one polarization direction $P_k$:
\begin{align}
    Q_{ikk}&=\dfrac{1}{2}\dfrac{\partial^2 x_{ii}}{\partial P_{k}^2} \\
M_{ikk}&=\dfrac{1}{2}\dfrac{\partial^2 x_{ii}}{\partial E_{k}^2}\quad , =\dfrac{1}{2}\dfrac{\partial^2 x_{ii}}{\partial P_{k}^2}\left[\dfrac{\partial P_{k}}{\partial E_{k}}\right]^2\label{eq:M}\quad .
\end{align}

The experimental reports of electrostrictive properties in literature in the past have mostly been relying on the $Q$-tensor, which was found to be relatively constant in the vicinity of ferroelectric phase transitions \cite{uwe1975, Li2014}. However, recently  Tanner \textit{et al} [\onlinecite{tanner}] have pointed out that the electrostrictive tensor relevant for mechanical applications is actually $M$ rather than $Q$ -- the tensor that links the strain to the applied electric field. This tensor diverges at the ferroelectric phase transitions since it is proportional to the square of the dielectric susceptibility $\chi_{kk}=\epsilon_{kk}-1$:
\begin{equation}
    M_{ikk}=Q_{ikk}\chi_{kk}^2 \quad .
\end{equation}
This relationship can be obtained directly from equation \eqref{eq:M}.
Quantitatively, our calculated values of the $Q$ tensor are in excellent agreement with the zero-temperature values for KTaO$_3$ calculated by Tanner \textit{et al} \cite{tanner}.
We find that the coupling of the polarization to the strain ($Q$) remains constant or decreases slightly with temperature (See Fig \ref{fig:electrostriction}a), which is consistent with experimental observations \cite{uwe1975}. The coupling between strain and applied electric field (M) increases sharply at low temperature (See Fig. \ref{fig:electrostriction}b) due to its direct dependence to the square of the dielectric constant. The values obtained for $M$ at temperatures below $\approx$100~K are of the same order of magnitude as the highest performing relaxor ferroelectrics \cite{yu2022}.

\section{Conclusion}
In this paper, we have shown that using a combination of quantum self-consistent lattice dynamics (QSCAILD) and machine-learning effective potentials (MLIPs) based on DFT calculations allows to explore accurately and finely the temperature dependence of physical properties of quantum paraelectric materials.
 Using KTaO$_3$ as an example, we have shown that our calculations are in excellent agreement with experiment, correctly reproducing the flattening out of the phonon frequency and of the dielectric constant at temperatures below 25\ K, which is a key feature of the quantum paraelectric state.
 
 We further explored the stability limits of the quantum paraelectric phase as a function of volume, epitaxial strain and temperature, and found that the phonon frequencies and dielectric constants are highly susceptible to small changes in lattice parameters even at room temperature, enabling the possibility of tuning the dielectric constant by the application of epitaxial strain.
Last but not least, it allowed us, for the first time, to calculate electrostrictive tensors as a function of temperature. Our theoretical calculations confirm that the strain response as a function of polarization remains rather stable over a large temperature range, in line with earlier experiments. Due to the direct relationship between the dielectric constant and the electrostrictive tensor, this implies that the electrostrictive response in KTaO$_3$ can be tuned by epitaxial strain, making it possible to obtain a giant electrostrostrictor at room temperature.
\\

\section{Acknowledgments}
This work was supported by the Swiss National Science Foundation under project No. P2EZP2\_191872 and by Institut Carnot under project PREDICT.

\bibliography{bib}

\end{document}